\documentstyle[11pt,newpasp,twoside,epsf]{article}
\input psfig.tex

\def\edcomment#1{\iffalse\marginpar{\raggedright\sl#1\/}\else\relax\fi}
\reversemarginpar

\def\apj{{ApJ}}
\def\apjs{{ApJS}}
\def\asec{$^{\prime\prime}$}

\def\etal{{et al. }}

\def\lamb{$\lambda$}
\def\lax{{$\mathrel{\hbox{\rlap{\hbox{\lower4pt\hbox{$\sim$}}}\hbox{$<$}}}$}}
\def\gax{{$\mathrel{\hbox{\rlap{\hbox{\lower4pt\hbox{$\sim$}}}\hbox{$>$}}}$}}
\def\simlt{\lower.5ex\hbox{$\; \buildrel < \over \sim \;$}}
\def\simgt{\lower.5ex\hbox{$\; \buildrel > \over \sim \;$}}

\def\nat{{Nature}}
\def\pasp{{PASP}}

\def\percm2{cm$^{-2}$}

\begin{document}
\title{Nonstandard Central Engines in Nearby Galaxies}
 \author{Luis C. Ho}
\affil{The Observatories of the Carnegie Institution of Washington, 813 Santa 
Barbara St., Pasadena, CA 91101, U.S.A.}

\begin{abstract}
We argue that nearby galaxy nuclei contain massive black holes
that are fueled by low radiative efficiency accretion flows.
\end{abstract}

\vspace*{-0.5cm}
\section{Nuclear Activity in Nearby Galaxies}

A significant fraction of local galaxies exhibit signs of nuclear activity in 
the form of emission-line nuclei classified as Seyferts or LINERs.  According 
to the Palomar spectroscopic survey of nearby galaxies, 43\% of all northern 
galaxies brighter than $B_T = 12.5$ mag are active (Ho, Filippenko, \& Sargent 
1997), albeit at a level substantially weaker than traditionally studied AGNs 
such as classical Seyfert nuclei and quasars.  The sheer abundance of 
low-luminosity AGNs (LLAGNs), especially LINERs which make up two-thirds of the 
population, compels us to give them proper attention when considering AGN 
issues of a statistical nature. 

We do not yet have a full understanding of the physical origin of LINERs.  
Particularly thorny are the narrow-line objects (type~2 LINERs and transition 
objects), where in some cases the AGN signature can be either ambiguous 
or absent (Ho 2001a; Barth 2001).  On the other hand, the affiliation of 
type~1 objects (those with detectable broad emission lines) with nonstellar 
processes seems quite secure.  This contribution highlights some of the salient 
features of type~1 LLAGNs and implications we might draw 
concerning the nature of their central engines.  

\vspace*{-0.3cm}
\section{Notable Characteristics of Low-luminosity AGNs}

We begin by listing some of the observational properties that are unique to 
LLAGNs.  When observing these sources, it should be noted that one must 
exercise caution to obtain {\it nuclear}\ fluxes --- the quantities most 
pertinent to the AGN and most analogous to observations of quasars.  In 
general the central source is sufficiently weak that it is completely 
overwhelmed by emission from the host galaxy.  This applies to virtually all 
wavelengths.  In order to reliably quantify the nuclear fluxes, one needs 
observations with good sensitivity, and more crucially, high angular 
resolution (generally \lax 1\asec).

\vspace*{0.1cm}
\noindent
{\it (1) Accretion luminosities and Eddington ratios.}\  LLAGNs are not
only intrinsically faint, but more importantly, their accretion luminosities
are low relative to their Eddington luminosities.  Ho (2002)
used the nuclear X-ray luminosities of a large sample of nearby galaxies
to estimate their nuclear bolometric luminosities. Figure~1{\it a}\ shows
that $L_{\rm bol}$ is generally higher in Seyferts 

\begin{figure}
\vbox{
\hbox{
\hskip 0truein
\psfig{file=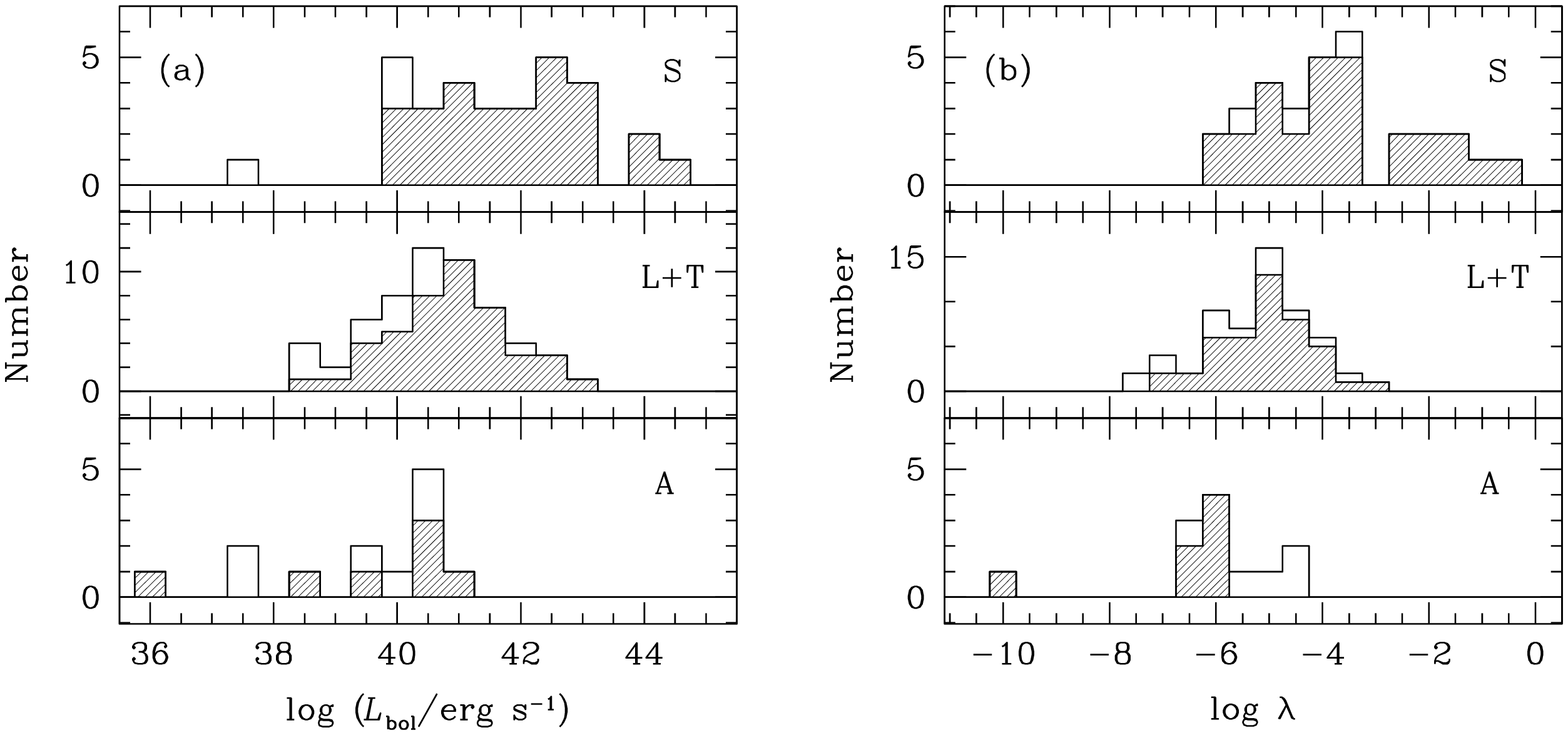,height=2.5truein,angle=270}
}
}
\vskip +0.7cm
\noindent{Fig. 1.} Distribution of ({\it a}) nuclear bolometric luminosities 
and ({\it b}) Eddington ratios $\lambda\,\equiv\,L_{\rm bol}/L_{\rm Edd}$ for
Seyferts (S), LINERs and transition objects (L+T), and absorption-line nuclei
(A).  Open histograms denote upper limits.  Adapted from Ho (2002).
\end{figure}

\vspace*{-0.9cm}
\noindent
than in LINERs,
which themselves are more luminous than absorption-line nuclei (objects with
no nuclear optical emission lines).   Using the empirical relationship between
black hole mass and bulge stellar velocity dispersion (Gebhardt et al. 2000;
Ferrarese \& Merritt 2000), one can obtain $L_{\rm Edd}$, and hence $\lambda$,
which is a function of $\dot{M}/\dot{M}_{\rm Edd}$.  Although the overlap is
considerable, $\lambda$ decreases systematically from Seyferts to LINERs
to absorption-line nuclei (Fig.~1{\it b}).  Note, in particular, that
{\it all}\ LINERs are characterized by \lamb\ \lax\ $10^{-3}$ and most have 
\lamb\ \lax\ $10^{-4}$. (It should be remarked that not all LLAGNs necessarily 
have low $\lambda$.  A good example is NGC 4395, one of the lowest luminosity 
AGNs known; it has $\lambda$ \gax\ $2\times 10^{-3}$ according to Moran et al. 
1999.)

\vspace*{0.1cm}
\noindent
{\it (2) Spectral energy distributions (SEDs).}\   With few 
exceptions, the SEDs of LLAGNs lack the optical--UV ``big blue bump,'' a feature
conspicuous in unobscured high-luminosity AGNs that is attributed to 
thermal emission from an optically thick, geometrically thin accretion disk. 
This unusual property was first quantified systematically by Ho (1999), and 
Ho et al. (2000) noted that it is also present in AGNs with double-peaked 
broad emission lines.  Figure~2 shows the full SEDs of the objects discussed by 
Ho et al. (2000), updated with NGC 4579, which Barth et al. (2001) recently 
discovered to belong to the same class.  Another attribute of these SEDs is 
that they are generically ``radio loud,'' defined here by the convention that 
the radio-to-optical luminosity ratio $R$ exceeds a value of 10.  In fact, 
radio loudness seems to be a property common to essentially {\it all}\ nearby 
weakly active nuclei (Ho 2001b) and a substantial fraction of Seyfert nuclei 
(Ho \& Peng 2001).  Moreover, the degree of radio loudness evidently changes 
systematically with accretion rate; $R$ increases with decreasing 
$\lambda\,\equiv\,L_{\rm bol}/L_{\rm Edd}$ (Fig.~3).

\vspace*{0.1cm}
\noindent
{\it (3) Structure of accretion disk.}\  The above-mentioned peculiar SEDs of 
LLAGNs cannot be readily accommodated by the predicted spectra of canonical
optically

\begin{figure}
\vbox{
\hbox{
\hskip 0.0truein
\psfig{file=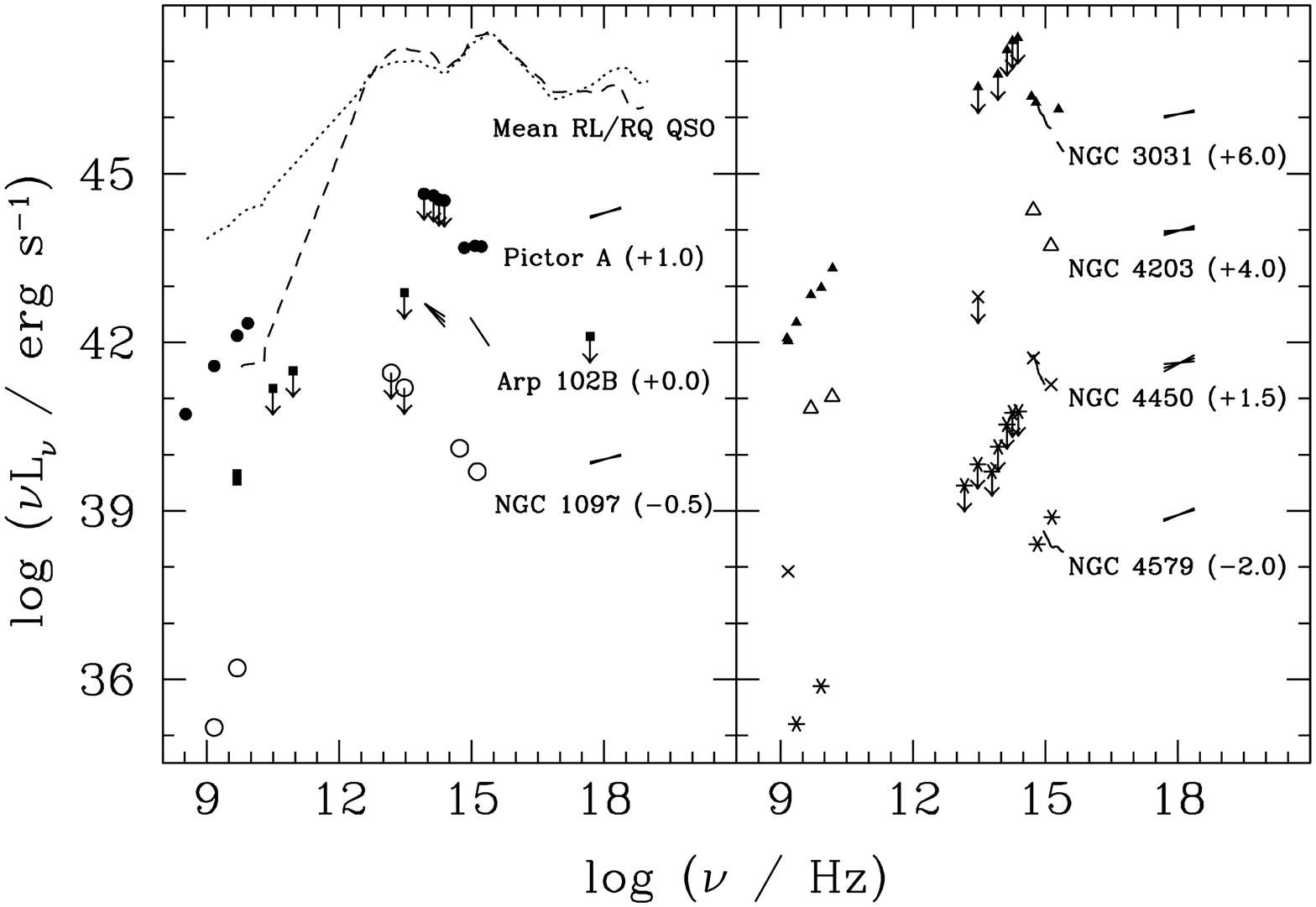,height=2.9truein,angle=270}
}
}
\vskip +1.2cm
\noindent{Fig. 2.} Nuclear SEDs of objects with double-peaked broad emission
lines.  Adapted from Ho et al. (2002).
\end{figure}

\vspace*{-0.7cm}
\noindent
thick, geometrically thin accretion disks.  The few cases that have
been modeled in detail need to invoke a central structure consisting of a
quasi-spherical, optically thin advection-dominated accretion flow (ADAF; see
reviews by Narayan, Quataert, \& Mahadevan 1998 and Quataert 2001), a concept 
similar to what Rees et al. (1982) called an ``ion-supported torus.''  ADAFs 
are thought to arise when the mass accretion rate drops below a critical 
threshold of $\dot{M}\,\approx\,10^{-2} \dot{M}_{\rm Edd}$.  In addition to a 
puffed-up inner hot corona, the SED fits of some objects (Lasota et al. 1996; 
Quataert et al. 1999; Lu \& Wang 2000) require a component from an outer thin 
disk.  The inner edge of the truncated disk, or the transition radius between 
the inner ADAF and outer thin disk, lies at 
$r\,\approx$ few $\times\,(10-100) R_S$, $R_S$ being the Schwarzschild radius.  

\vspace*{0.1cm}
\noindent
{\it (4) Double-peaked broad emission lines.}\  The ADAF + truncated disk 
cartoon for the central structure (see Fig.~4) is further implicated from the 
recent frequent detection of double-peaked broad emission lines in 
nearby LLAGNs (Ho et al. 2000, and references therein).  

\vspace*{0.1cm}
\noindent
{\it (5) X-ray Fe~K$\alpha$ line.}\  The X-ray spectra of LLAGNs generally 
require a hard power-law component with photon indices $\Gamma\,\approx\,
1.7-1.9$, but the relativistically broadened Fe~K$\alpha$ line at 6.4 keV 
is either extremely weak or absent (Terashima et al. 2001).  The Fe~K is 
thought to arise from fluorescence off of cold material near the center 
that subtends a large solid angle, commonly interpreted to be the standard
thin accretion disk.  The absence of this feature in LLAGNs is consistent
with the accretion-disk structure suggested by (3) and (4) above.

\vspace*{0.1cm}
\noindent
{\it (6) Low-ionization state.}\ As mentioned in \S~1, the majority of nearby 
LLAGNs are classified as LINERs, objects with lower ionization state than 
Seyferts.  It is notable that radio galaxies with double-peaked broad lines 
(e.g., Arp 102B, 

\begin{figure}
\vbox{
\hbox{\vsize 2.7in
\hskip -0.1truein
\psfig{file=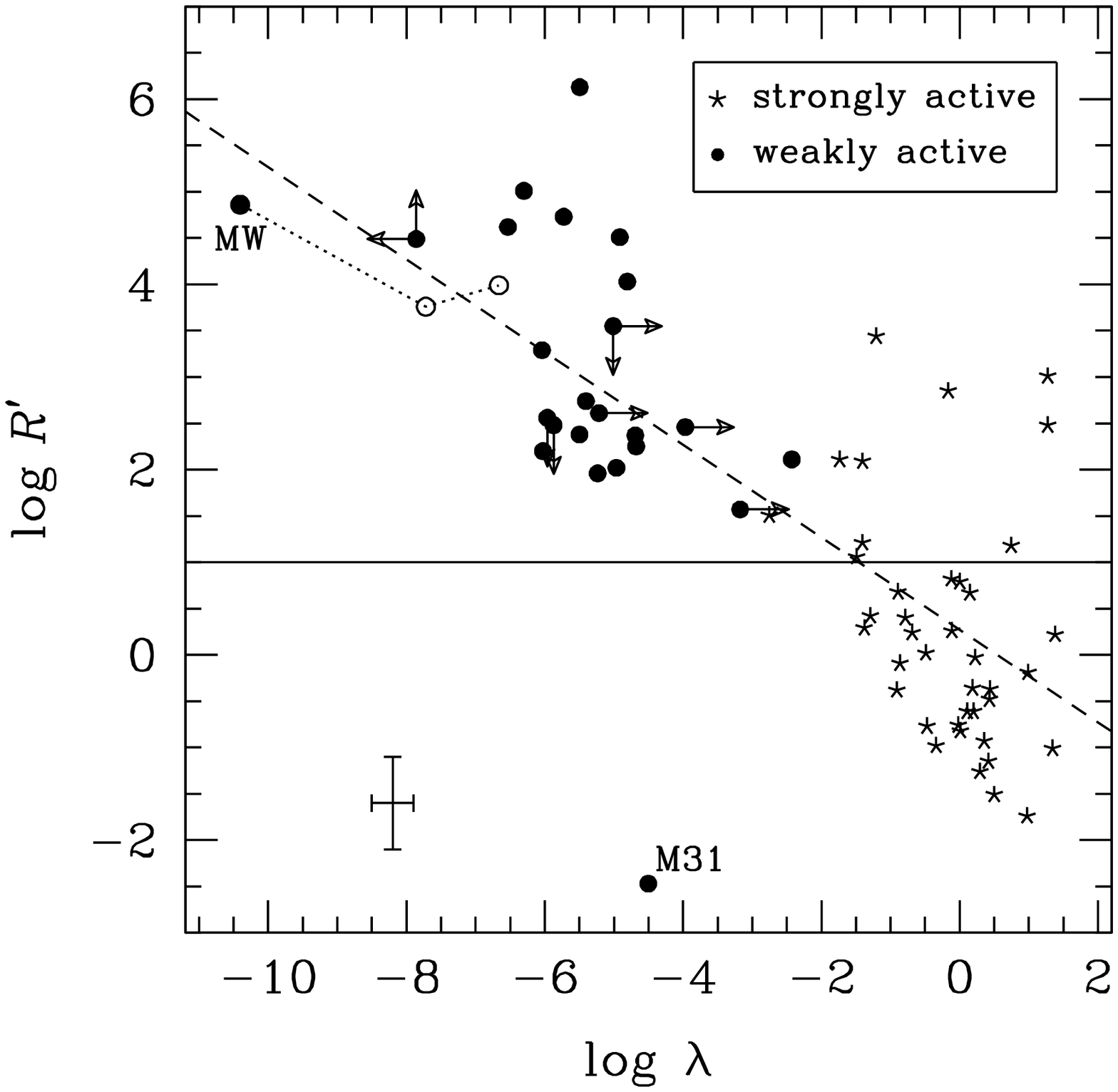,height=3.2truein,angle=0}
}
\vspace*{-2.4in}
\hskip +3.4truein
\vbox{\hsize 1.8in
\noindent{Fig. 3.} Distribution of the nuclear radio-to-optical luminosity
ratio $R^{\prime}$ vs.  $\lambda\,\equiv\,L_{\rm bol}/L_{\rm Edd}$. The
{\it solid line}\ marks the formal division between radio-loud and radio-quiet
objects, $R^{\prime}$ = 10.  The {\it dashed line}\ is the best-fitting linear
regression line.  Adapted from Ho (2001b).
}
}
\end{figure}

\vspace*{0.2cm}
\noindent
3C 390.3, Pictor A) also typically have LINER-like 
narrow-line ratios (e.g., Eracleous \& Halpern 1994).


\vspace*{-0.2cm}
\section{A Unifying Physical Picture}
\vspace*{-0.1cm}

We propose that the above set of characteristics of LLAGNs can be accommodated 
within the theoretical framework of an ADAF, or some closely related variant 
thereof (Narayan et al. 1998; Quataert 2001).  The picture we envision is 
depicted in Figure~4: a hot, ion-supported torus or ADAF exists within some 
transition radius, exterior to which it illuminates a thin disk.  
ADAFs have radiative efficiencies much less than the canonical 10\% for thin 
disks because they cannot cool effectively, and so they naturally produce low 
luminosities.  Consistent with the models, the observed Eddington luminosity 
ratios of LINERs invariably lie below the critical threshold of $\lambda$ \lax\ 
$10^{-2}$.  Several emission mechanisms (cyclo-synchrotron, inverse Compton 
scattering, and bremsstrahlung) give rise to the broad-band spectrum from radio 
to X-ray energies.  With an absent or truncated thin disk, however, the 
accretion structure generates little or no emission from the traditional 
optical--UV big blue bump.  Emission in this band instead comes from inverse 
Compton scattering of cyclo-synchrotron photons, and its strength increases 
sensitively with rising $\dot{M}/\dot{M}_{\rm Edd}$ (see e.g., Fig.~1 of 
Mahadevan 1997).  The radio component, on the other hand, is persistently 
prominent because cyclo-synchrotron emission contributes significantly to the 
cooling budget of an ADAF.  Both of these features qualitatively account for 
the peculiar shape of the SEDs, their generic ``radio loudness,'' and the 
inverse correlation between $R$ and $\lambda$ (Fig.~3).  In detail, the steeply 
falling NIR--UV spectrum ($f_{\nu}\,\propto\,\nu^{-\alpha}$, $\alpha = 1.5-2.5$)
observed in many objects (Ho 1999; Fig.~2) suggests an extra component in 
additional to an ADAF.  This emission plausibly comes from an outer thin disk.  
With an inner radius located at $r\,\approx$ few $\times\,(10-100) R_S$, the 
disk is relatively cool: the big blue bump shifts to a ``big red bump,'' and the 
NIR--UV tail is equivalent to the soft X-ray excess component 

\begin{figure}
\vbox{
\hbox{
\hskip 0.4truein
\psfig{file=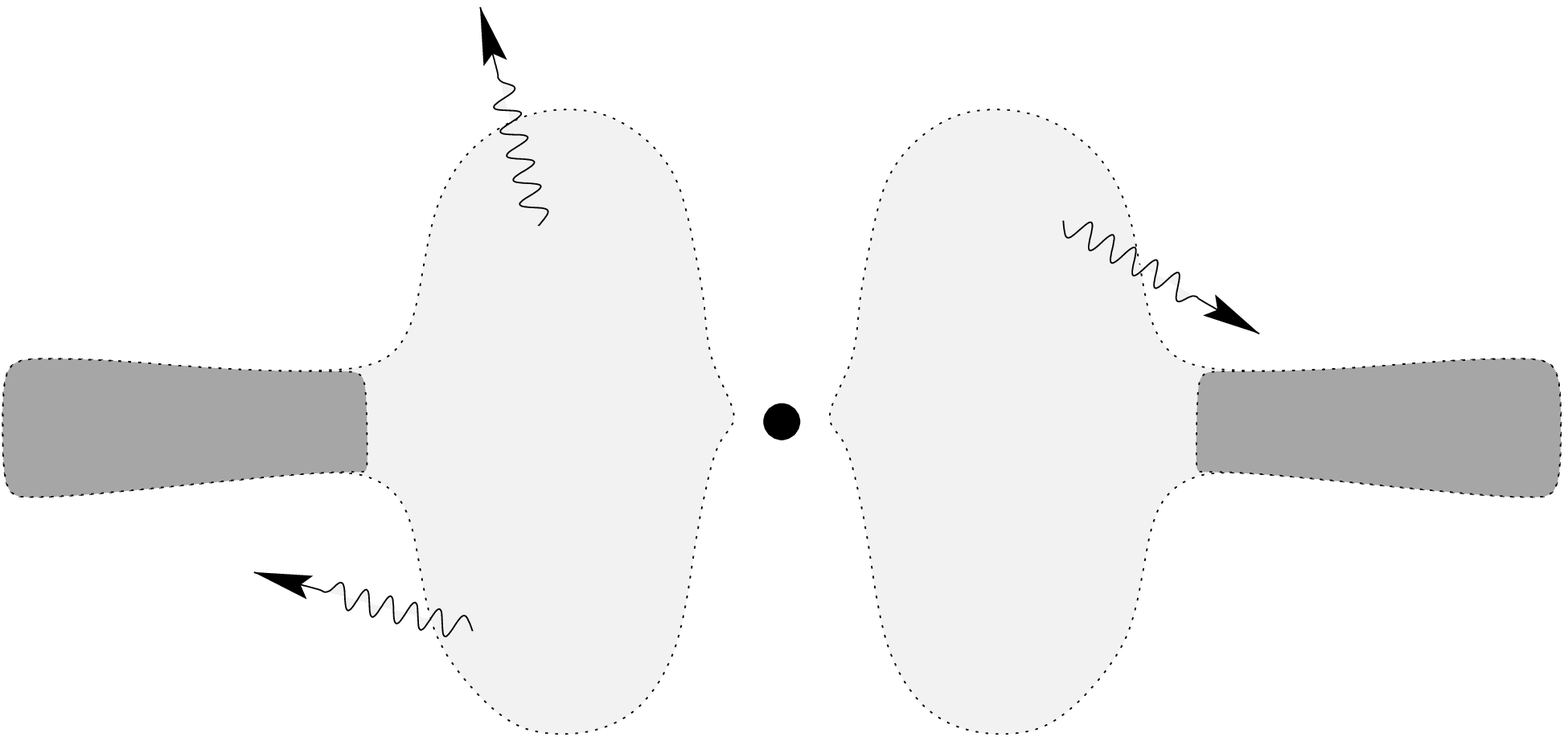,height=2.1truein,angle=0}
}
}
\vskip +0.5cm
\noindent{Fig. 4.} A cartoon depicting the structure of the accretion flow
surrounding weakly active massive black holes.  An inner ADAF (ion-supported
torus) irradiates an outer thin disk.  Taken from Shields et al. (2001).
\end{figure}

\vspace*{-0.3cm}
\noindent
of high-luminosity 
AGNs.  This very same outer disk has been invoked to produce the double-peaked 
broad Balmer lines (e.g., Chen, Halpern, \& Filippenko 1989).  The outer thin 
disk, even when present, subtends too small a solid angle with respect to the 
primary X-rays from the ADAF to generate strong Fe~K$\alpha$ emission.  

Lastly, we note that low-ionization spectra may emerge quite naturally in the 
scenario suggested here.  In the context of AGN photoionization models, it is 
well known that LINER-like spectra can be produced largely by lowering the 
``ionization parameter'' $U$, typically by factors of $10-100$ below those in 
Seyferts (e.g., Halpern \& Steiner 1983; Ferland \& Netzer 1983; Ho, 
Filippenko, \& Sargent 1993).  Figure~1{\it a}\ shows that the nuclear 
luminosities of LINERs indeed are at least a factor of 10 lower than in 
Seyferts.  In fact, the weakness of the UV emission in the SEDs of LINERs 
suggests that the ionizing luminosity should be reduced by an even larger 
factor.  Two other effects may be important in boosting the low-ionization 
lines.  All else being equal, hardening the ionizing spectrum (by removing the 
big blue bump) in photoionization calculations creates a deeper partially 
ionized zone from which low-ionization transitions, especially [O~I] \lamb\lamb 
6300, 6363, are created.  Because of the prominence of the radio spectrum, 
cosmic-ray heating of the line-emitting gas by the mildly relativistic 
electrons in the ADAF may be nonnegligible; one effect of this process is again 
to enhance the low-ionization lines (Ferland \& Mushotzky 1984).

\vspace*{0.1cm}

\acknowledgments{
L.C.H. acknowledges financial support through NASA grants from the Space 
Telescope Science Institute (operated by AURA, Inc., under NASA contract 
NAS5-26555).}


\vspace*{0.0cm}
\begin{references}
\vspace*{-0.1cm}


\reference 
Barth, A.~J. 2001, in Issues in Unification of AGNs, ed. R. Maiolino, A. 
Marconi, \& N. Nagar (San Francisco: ASP), in press

\reference 
Barth, A.~J., Ho, L.~C., Filippenko, A.~V., Rix, H.-W., \& Sargent, W.~L.~W.
2001, \apj, 546, 205

\reference 
Chen, K., Halpern, J.~P., \& Filippenko, A.~V. 1989, \apj, 339, 742

\reference 
Eracleous, M., \& Halpern, J.~P. 1994, \apjs, 90, 1

\reference 
Ferland, G. J., \& Mushotzky, R.~F. 1984, \apj, 286, 42

\reference 
Ferland, G. J., \& Netzer, H. 1983, \apj, 264, 105

\reference 
Ferrarese, L., \& Merritt, D. 2000, \apj, 539, L9

\reference 
Gebhardt, K., \etal 2000, \apj, 539, L13

\reference 
Halpern, J.~P., \& Steiner, J.~E. 1983, \apj, 269, L37

\reference 
Ho, L.~C. 1999, \apj, 516, 672

\reference 
------. 2001a, in IAU Colloq. 184, AGN Surveys, ed. R.~F. Green, E.~Ye.
Khachikian, \& D.~B. Sanders (San Francisco: ASP), in press

\reference 
------. 2001b, ApJ, in press

\reference 
------. 2002, in preparation

\reference 
Ho, L.~C., et al. 2001, \apj, 549, L51

\reference 
------. 2002, in preparation

\reference 
Ho, L.~C., Filippenko, A.~V., \& Sargent, W.~L.~W. 1993, \apj, 417, 63

\reference 
------. 1997, \apj, 487, 568
 
\reference 
Ho, L.~C., \& Peng, C.~Y. 2001, \apj, 555, 650

\reference 
Ho, L.~C., Rudnick, G., Rix, H.-W., Shields, J.~C., McIntosh, D.~H.,
Filippenko, A.~V., Sargent, W.~L.~W., \& Eracleous, M. 2000, \apj, 541, 120

\reference 
Lasota, J.-P., Abramowicz, M.~A., Chen, X., Krolik, J., Narayan, R., \& Yi, I.
1996, \apj, 462, 142

\reference 
Lu, Y., \& Wang, T. 2000, \apj, 537, L103

\reference 
Mahadevan, R. 1997, \apj, 477, 585

\reference 
Moran, E.~C., Filippenko, A.~V., Ho, L.~C., Shields, J.~C., Belloni, T.,
Comastri, A., Snowden, S.~L., \& Sramek, R.~A. 1999, \pasp, 111, 801

\reference 
Narayan, R., Mahadevan, R., \& Quataert, E. 1998, in The Theory of Black Hole
Accretion Discs, ed.  M. A. Abramowicz, G. Bj\"{o}rnsson, \& J. E. Pringle
(Cambridge: Cambridge Univ. Press), 148


\reference 
Quataert, E. 2001, in Probing the Physics of Active Galactic 
Nuclei by Multiwavelength Monitoring, ed. B.~M. Peterson, R.~S. Polidan, \& 
R.~W. Pogge (San Francisco: ASP), 71

\reference 
Quataert, E., Di Matteo, T., Narayan, R., \& Ho, L.~C. 1999, \apj, 525, L89

\reference 
Rees, M.~J., Begelman, M.~C., Blandford, R.~D., \& Phinney, E.~S. 1982,
\nat, 295, 17

\reference 
Shields, J.~C., et al. 2001, in Probing the Physics of Active Galactic 
Nuclei by Multiwavelength Monitoring, ed. B.~M. Peterson, R.~S. Polidan, \& 
R.~W. Pogge (San Francisco: ASP), 327

\reference 
Terashima, Y., Iyomoto, N., Ho, L.~C., \& Ptak, A.~F. 2001, \apjs, in press

\end{references}
\end{document}